\documentstyle[11pt,newpasp,twoside,epsf]{article}
\def\edcomment#1{\iffalse\marginpar{\raggedright\sl#1\/}\else\relax\fi}
\reversemarginpar

\begin{document}
\title{Active Galaxies at Milliarcsecond Resolution in the NOAO
       Deep Wide-Field Survey}

\author{J.~M. Wrobel, T.~A. Rector, G.~B. Taylor, \& S.~T. Myers}
\affil{National Radio Astronomy Observatory, Socorro, NM, U.S.A.}
\author{C.~D. Fassnacht}
\affil{University of California, Davis, CA, U.S.A.}

\begin{abstract}
We are using the NRAO VLBA at 5.0~GHz to image about 200 FIRST sources
stronger than 10~mJy at 1.4~GHz in the NDWFS.
\end{abstract}

\section{Motivation}

Traditional continuum surveys using Very Long Baseline Interferometry
(VLBI) have had limited astrophysical impact, because they suffer from
a paucity of redshifts and from biases introduced by targeting bright
and, often, flat-spectrum radio sources.  To overcome these
limitations, we are using the NRAO Very Long Baseline Array (VLBA) at
5.0~GHz to image about 200 compact FIRST sources stronger than 10~mJy
at 1.4~GHz (White et al.\ 1997) in the NOAO Deep Wide-Field Survey
(NDWFS).  Optical identifications to 26 magnitudes (mag) are becoming
available from the NDWFS (Jannuzi \& Dey 1999) and spectroscopic
follow-up has commenced for many of the FIRST sources.

\section{Observations}

Our VLBA (Napier et al.\ 1994) survey at 5.0~GHz uses phase
referencing to reach about 100 FIRST sources in each of the Bo\"otes
and Cetus fields of the NDWFS.  Each compact FIRST source was selected
to have a largest angular size less than 5$\arcsec$.  The error
ellipse for the {\em a priori\/} FIRST position typically had a FWHM
of 850 milliarcseconds (4$\sigma$).  The phase referencing, in the
nodding style, involved a switching time of 5~minutes and switching
angles of 2$\fdg$5 or less.  Each FIRST source was observed with a
total bandwidth of 64~MHz during six 80-second snapshots spread over
time to enhance coverage in the $(u,v)\/$ plane.  Our VLBA survey is
complete in the Bo\"otes field and in progress in the Cetus field,
with each field covering 9 square degrees.

\section{Results}

In the Bo\"otes field, about one source in three was detected with the
VLBA as stronger than 1.5-2.5~mJy (6$\sigma$) at 2-milliarcsecond
resolution.  Most VLBA detections were unresolved but four apparent
doubles were discovered.  These doubles are unlikely to be
gravitational millilenses, as the upper limit at 95\%-confidence to
the expected lensing rate is about 1 lens per 430 sources (Wilkinson
et al.\ 2001).  Rather, these doubles might be Compact Symmetric
Objects (CSOs), young and rare systems offering insights into
evolutionary models for radio galaxies and strong tests of unified
schemes (eg, Taylor et al.\ 2000).  Follow-up VLBA observations are
planned to test whether or not these doubles are indeed CSOs.

Optical identifications to 26 mag (5$\sigma$) at 2$\arcsec$ resolution
are becoming available for the Bo\"otes field of the NDWFS.  At these
depths we expect that almost all FIRST sources will be identified,
mostly with radio galaxies and quasars (Waddington et al.\ 2000).  By
contrast, less than a third of FIRST sources are identified to 22 mag
in the Sloan Digital Sky Survey (Ivezi\'c et al.\ 2002).
Identifications are an essential prerequisite for redshifts.  NDWFS
identifications from early release data covering 1.4 square degrees
are described in the figure for four VLBA detections.

Our VLBA images either locate the active nuclei within the optical
hosts, or impose upper limits on emission from the active nuclei.  We
will constrain the spectral energy distributions of the active nuclei
by combining the VLBA data with the photometric data from the NDWFS,
from Chandra (C.\ Jones \& S.\ Murray), and from SIRTF (P.\
Eisenhardt), as those data become available.  The NDWFS and SIRTF data
will also serve to constrain the galaxy cluster environments of the
active nuclei.

The stronger VLBA detections can serve as in-beam phase calibrators
for deep, wide-field VLBI imaging of the microJy sky in the Bo\"otes
field (de Vries et al.\ 2002; Morganti \& Garrett 2002).  Results from
a 24-hour pilot study, using the VLBA and the NRAO Green Bank
Telescope at 1.4~GHz, are presented in these Proceedings (Garrett,
Wrobel, \& Morganti 2004).

\acknowledgments This work made use of images provided by the NDWFS,
which is supported by NOAO.  NOAO is operated by AURA, Inc., under a
cooperative agreement with the NSF.  NRAO is a facility of the NSF
operated under cooperative agreement by AUI.

\begin{figure}
\plotone{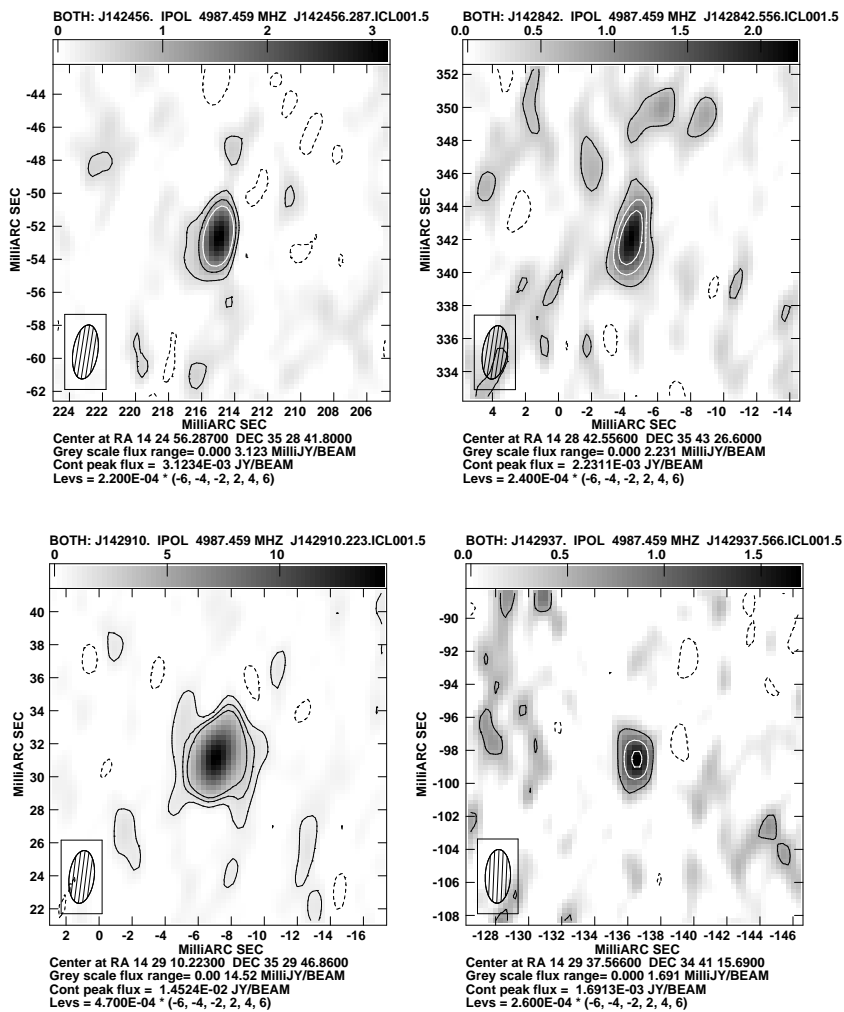}
\caption{Images of Stokes I emission at 5.0~GHz for four VLBA
detections with NDWFS identifications.  Boxed ellipse shows the
Gaussian restoring beam at FWHM.  Contours are at $\pm$2, $\pm$4, and
$\pm$6 times the quoted rms noise levels.  Angular offsets are
relative to the quoted FIRST positions.
{\em Top    left:}  J142456.287$+$352841.80, $I\sim$~19.8~mag quasar.
{\em Top    right:} J142842.556$+$354326.60, $I\sim$~19.2~mag galaxy.
{\em Bottom left:}  J142910.223$+$352946.86, $I\sim$~18.4~mag quasar.
{\em Bottom right:} J142937.566$+$344115.69, $I\sim$~21.5~mag galaxy.}
\end{figure}

\end{document}